\RequirePackage{fix-cm}
\documentclass{svjour3} 
\smartqed  
\usepackage{graphicx}
\usepackage{amsmath}
\usepackage{natbib}
\usepackage{amsfonts}
\usepackage{amssymb}
\usepackage{epstopdf}
\usepackage{etoolbox}
\usepackage[misc,geometry]{ifsym} 
\usepackage{lipsum}
\begin{document}

\title{Modeling the impact of rain on population exposed to air pollution}



\author{  Sandeep Sharma    \and  Nitu Kumari \
        }


\institute{Sandeep Sharma \and Nitu Kumari \at
          School of Basic Sciences\\
          Indian Institute Of Technology, Mandi\\
           Mandi, Himachal Pradesh-175001, India \\
          \email{sampark81@gmail.com}             \\
            \and 
             Nitu Kumari (\Letter) \at
            \email{nitu@iitmandi.ac.in}
}
\date{Received: date / Accepted: date}

\maketitle

\begin{abstract}
Environmental pollution, comprising of air, water and soil have emerged as a serious problem in past two decades. The air pollution is caused by contamination of air due to various natural and anthropogenic activities. The growing air pollution has diverse adverse effects on human health and other living species. However, a significant reduction in the concentration of air pollutants has been observed during the rainy season. Recently, a number of studies have been performed to understand the mechanism of removal of air pollutants due to the rain. These studies have found that rain is helpful in removing many air pollutants from the environment. In this paper, we proposed a mathematical model to investigate the role of rain in removal of air pollutants and its subsequent impacts on human population.
\keywords{Air pollutants \and PM \hspace{-0.2cm}\begin{tiny}
10
\end{tiny} \and PM \hspace{-0.2cm}\begin{tiny}
2.5
\end{tiny} \and Precipitation \and Mathematical Model \and Stability}
\end{abstract}
\section{Introduction}
\label{intro}
Since the beginning of industrial revolution, air pollution has plagued the industrialized cities of world. There is a significant increase in the concentration of air pollutants in the atmosphere. An \textit{air pollutant}, in simple words, can be defined as an unwanted substance having harmful effects on humans, animals, and climate \citep{kampa2008human}. Air pollutants can be categorized as primary and secondary pollutants. Pollutants emitted directly from sources like industrial discharge, vehicular exhaust, and household discharge form the class of primary pollutants. The primary pollutants generate secondary pollutants through chemical transformation \citep{naresh2006modeling}.
 
Although, many natural activities (e.g. volcanoes, forest fire etc.) contribute to the air pollution but anthropogenic activities are the leading cause of damage to the environment. A variety of air pollutants consisting of hydrocarbons, poisonous gases (e.g CO, SO\begin{tiny}2\end{tiny}, NO etc.), metals (Lead, Mercury etc.) and particulate matters (PM\begin{tiny}2.5\end{tiny}, PM\begin{tiny}10\end{tiny} etc.) can escape from industry, traffic and other everyday activities. The hazardous effect of air pollution is a well-established fact since antiquity. Recent studies confirm a direct correlation between sleazy air quality and acute lower respiratory infections. An exacerbation of premature human mortality caused by lung cancer and cardiopulmonary disease has also been observed in different epidemiological studies \citep{pope2002lung,pope2004cardiovascular,nawrot2006environmental,miller2007long}. 
A rapid increase in air pollution results in poor quality of air, which in turn has many adverse impacts on human health. As per the WHO, in 2012 around 7 million people died due to air pollution. This means that almost every eighth individual succumb to air pollution. This finding is more than twice of the previous estimates and places air pollution at the top among all the world's environmental health risks \citep{whoair}. \cite{kampa2008human} performed a detailed review on different categories of air pollutants and their respective health effects. 
 
The particulate matters, emitted from combustion of fossil fuels, enter into the human body through inhalation. The exposure to particulate matters (PMs) for  short and long period of time resulted in an increased risk of CV (cardiovascular) mortality and morbidity. Apart from the respiratory problems, these pollutants can also cause hematological  problems and cancer. Even inhalation of PM\begin{tiny}{2.5}\end{tiny} over shorter periods can trigger cardiac ischemia, myocardial infarctions, heart failure, strokes, arrhythmias and sudden death. However, the long-term exposures may also increase the risk of developing chronic cardiovascular diseases \citep{brook2008cardiovascular}. Exposure to combustion-related fine particles for a long duration increased the chance of lung cancer and caused cardiopulmonary mortality  \citep{pope2002lung}. Continuous exposure to fine particulates enhance the chances of specific cardiovascular disease mortality through mechanisms which include systemic and pulmonary inflammation,  altered cardiac autonomic function and accelerated atherosclerosis  \citep{pope2004cardiovascular}. \cite{tager2005chronic} conducted a study on University of California, Berkeley students to assess the effects of lifetime exposure to O\begin{tiny}3\end{tiny}, nitrogen dioxide (NO\begin{tiny}2\end{tiny}) and small particulate matter (PM\begin{tiny}10\end{tiny}). The study concluded that exposure to ozone reduces lung function. \cite{nawahda2012evaluation} performed a study to assess the premature mortality risk caused by exposure to particulate matter PM\begin{tiny}2.5\end{tiny} and Ozone in East Asia. The study found that PM\begin{tiny}2.5\end{tiny} has a higher negative impact as compared to ozone on individuals of an age group of 30 years and above.
   
Organic pollutants consisting of pesticides, dioxins, furans, and PCBs form a class of toxic chemicals. These chemicals remain in the food chain for a long time  and their effect increases on the population as it goes to the higher level of the food chain. Organic pollutants, in general, and dioxin, in particular, alters a variety of biological responses which includes including disruption of normal hormone signaling pathways, induction of cytochrome P-450 1A1 (CYP1A1), immunotoxicity, reproductive and developmental defects, wasting syndrome, liver damage, and cancer \citep{mandal2005dioxin}. It is also observed that chronic and subchronic non-cancer hazards and cancer risks are increased by exposure to hydrocarbons emitted from an NGD project in Colorado \citep{mckenzie2012human}.
 
 The earth crust naturally contains various heavy metals like lead, mercury, cadmium etc. Heavy metals are also generated due to combustion, waste water discharge, and manufacturing facilities. Although, low concentration of these heavy metals is an essential requirement of the human body but in the case of high concentration they become toxic and cause bioaccumulation. Sleep disorder, memory disturbances, fatigue and blurred vision are some of the adverse effects of high concentration of heavy metals in the human body. Exposure to heavy metals is responsible for emphysema, asthma, and lung cancer \citep{nawrot2006environmental}. 
 
 Among all the anthropogenic activities, combustion of fuels and industrial discharges are major sources of different types of air pollutants. Available literature suggests an increased contribution of traffic-related air pollution in poor air quality. Deforestation plays a significant role in the increase of air pollution. Urban air pollution is related to blood coagulation, inflammation, autonomic dysfunction and oxidative stress in healthy young populations, with O\begin{tiny}3\end{tiny} and sulfate as two major traffic-related pollutants causing such effects \citep{chuang2007effect}.
 \cite{laumbach2012respiratory} studied the hazardous effects of biomass fuels (BMF) and traffic-related air pollution (TRAP). The study finds a significant relationship between exposure to BMF and TRAP and diseases like asthma and tuberculosis.

 
Children are the most vulnerable group to the hazardous effects of ambient air pollution which begin even before the child is born \citep{salvi2007health}. \cite{van2015prenatal} performed a study on the impact of pollution on newborn systolic blood pressure (SBP) due to antenatal exposure to air pollution. The study concluded that exposures O\begin{tiny}3\end{tiny} in the later stage of pregnancy were negatively associated with SBP whereas particulate matters and BC were positively associated with newborn SBP.  \cite{tang2014air} studied the effects of exposure to polycyclic aromatic hydrocarbons (PAHs) on child growth and development before and after the shutdown of a coal-fired power plant in Tongliang, China.
 
It is easy to conclude, from above discussion that air pollutants have several hazardous effects on  living species and the environment. Hence, air pollution is a challenging problem which requires immediate attention. 

On the other hand, various studies have shown that rain has played a significant role in the removal of almost all types of air pollutants. A considerable decrement in the concentration of air pollutants has been observed during the rainy season. The atmosphere of the cities becomes clean after the rainy season. It has been observed in experimental studies, that different air pollutants are absorbed/trapped in raindrops and thus removed from the atmosphere by precipitation. The phenomena of precipitation scavenging  is important as it plays a crucial role in the removal of air pollutants from the environment. Hence, it is evident that the phenomenon of  absorption/impaction of pollutants plays a pivotal role in the  removal of air pollutants. In recent years, various mathematical and non-mathematical studies are performed to understand the effect of rain in the removal of air pollutants. 

\cite{hill1977model} developed a model  to study the washout of sulfur dioxide from the environment by rain. \cite{sharma1983atmospheric} performed an air quality study by measuring the level of different particulate matters in the environment of Kanpur city. The study concluded that concentration of particulate matters decreases considerably in monsoon season. \cite{pandey1992air} conducted an exhaustive study of various air pollutants present in the environment of Varanasi. The study observed that concentration of the air pollutants is highest in summer and lowest during the rainy season. \cite{ravindra2003variation} studied spatial patterns of different air pollutants, like O\begin{tiny}3\end{tiny}, NO\begin{tiny}2\end{tiny},   TSP and SO\begin{tiny}2\end{tiny} at Shahdara National Ambient Air Quality Monitoring station in Delhi (India). It is found that these spatial patterns are same before and during the rain. Further, It is also observed that concentrations of NO\begin{tiny}2\end{tiny}, TSP, and SO\begin{tiny}2\end{tiny} decreases significantly (40-45$\%$) after initial and subsequent rainfall, demonstrating the importance of rainfall in scavenging of air pollutants.
\cite{jolliet2005modeling} studied the impact of rain on long-range transport of organic substances in the air.  They used a new modeling approach to achieve this goal. The approach provides an explicit description of the role of substance parameters and precipitation. The mathematical model proposed by \cite{naresh2006modeling} demonstrates the role of rain in the removal of primary as well as secondary pollutants from the environment of an industrial city. It is observed that if primary pollutants are emitted instantaneously, then rain can completely remove the same from the environment. The primary, as well as secondary pollutants, are also washed out in the case of constant emission of primary pollutants.  \cite{naresh2007modeling}  used a deterministic mathematical model for investigate the role of precipitation in the removal of particulate matters and gaseous pollutants from the atmosphere of a city. The study found that complete removal of pollutants is possible if the rate of precipitation is large. \cite{shukla2008effect} studied a mathematical model comprising a system of five differential equations to study the scavenging of gaseous pollutants and particulate matters by rain from the environment of a city. Through the analysis of the model, it has been found that rain can remove the air pollutants from the environment. \cite{shukla2008modelling} used a mathematical model for understanding the role of cloud density in the removal of different air pollutants by rain. The study found that it is possible to remove air pollutants if cloud droplets and raindrops are formed at a sufficiently large rate. \cite{duhanyan2011below} performed a detailed review of frequently cited theories and parameterisations to demonstrate the role of rain in below-cloud scavenging. The study inspected effects of terminal velocity of raindrops, raindrop size distribution (RSD) and below-cloud scavenging coefficient for different air  pollutants including gaseous pollutants and particulate matters. \cite{yoo2014new} used routinely available air-monitored and meteorological data to investigate the washout effect of summertime rain on surface air pollutants (PM\begin{tiny}10\end{tiny}, CO, O\begin{tiny}3\end{tiny}, NO\begin{tiny}2\end{tiny} and SO\begin{tiny}2\end{tiny}). The study found statistically significant negative correlations between the concentrations of various air pollutants and rain intensity due to convection or washout.

From the above discussion, dangerous effects of air pollutants and their removal due to rain are apparent and sufficient literature is available on both these affairs. Despite this fact, no mathematical study is available to examine the impact of the removal of pollutants due to rain on the human population. Such study provides a clear insight of the phenomenon and is helpful to understand the role of various parameters involved in this whole process. In the present study, a new mathematical model is formulated to study the importance of rain in the removal of air pollutants and its positive impact on the human population. 

The manuscript is organized into four sections. The mathematical model is formulated in section (\ref{MM}). The results obtained through stability analysis and numerical simulation are discussed in section (\ref{RND}). The conclusion of the present work is presented in section (\ref{con}).

\section{Mathematical Model} 
\label{MM}
The environment of the urban area is highly contaminated with various air pollutants and has several adverse effects on human population. This alarming situation draws attention from every corner of the society. Several biological, statistical, experimental and mathematical studies have been performed on this line and are available in the literature (\cite{dubey2010model,dubey2000modelling,freedman1991models,hallam1983effects,zhien1990persistence,
hallam1984effects,huaping1991threshold} and references cited therein). These studies investigate various dimensions of the problem of air pollution and its effect on living species. However, a study on  simultaneous effect of rain and air pollutants on human population density have not been carried out.  

To accomplish this objective, in the current study a mathematical model consisting of the human population, air pollution, and rain is proposed and analyzed to understand the effect of rain on air pollution and human population.

The model has been formulated under the following assumptions \newline

1. There is  no migration of population from the habitat \citep{dubey2010model}. \newline

2. The growth rate of raindrops is constant \citep{naresh2006modeling,naresh2007modeling,shukla2008effect}. \newline

3. Air pollutants are emitted into the environment at a constant rate and the concentration of pollutants depletes due to some natural factors such as interaction with plant leaves, chemical hydrolysis, photosynthesis and biological transformation \citep{dubey2010model}.  \newline

4. Due to the different hazardous  effects of pollution, human population density decreases. The rate at which human population decreases assumed proportional to the product of the concentration of pollutants and the density of population \citep{dubey2010model}. \newline

5. The concentration of pollutants depletes due to absorption, uptake, and decomposition assumed  proportional to the product of the concentration of pollutants and the density of population \citep{dubey2010model}.  \\
\begin{equation}
\label{eq:1}
\frac{dN}{dt}= \Lambda - \beta N P - d N 
\end{equation}
\begin{equation}
\label{eq:2}
\frac{dP}{dt}= Q - \delta_0 P - \delta_1 P N - \delta_2 P R  
\end{equation}
\begin{equation}
\label{eq:3}
\frac{dR}{dt}= q - \alpha_0 R - \alpha P R
\end{equation}
In the above model system, N(t) is the density of human population, P(t) is the concentration of air pollutants and R(t) is the density of raindrops. The growth rate of human population is $\Lambda$ while d is the natural death rate. $\beta$ is the depletion rate of human population due to pollution. The  rate of emission of air pollutants into the environment is $Q$ and the growth rate of rain drops is $q$.  The natural depletion of pollutants also takes place at a constant rate $\delta_0$. $\delta_1$ and $\delta_2$ are the depletion rates of pollutants due to its uptake by human population and rain respectively. $\alpha_0$ is the natural depletion rate of rain drops. While $\alpha$ is the rate of depletion of raindrops due to air pollutants.  

The mathematical models available in the literature either addresses the issue of perilous effects of pollutants on the living species or the removal of air pollutant due to rain (see section (\ref{intro})). To the best of our knowledge, there is no mathematical model in the literature which exhorts the simultaneous impact of rain and pollution on the human population. The current study will certainly bridge this gap.

To perform the study, it is required to obtain the bounds on the dependent variables, i.e. $N, P$ and $R$. To achieve this task, the region of attraction \citep{freedman1985global} (without proof) is obtained as \newline

$\Omega = \lbrace (N,P,R): 0 \leq N \leq \frac{A}{d}, 0 \leq P \leq \frac{Q}{\delta_0}, 0 \leq R \leq \frac{q}{\alpha_0}\rbrace$ \newline

The model system possesses only one non-negative equilibrium point $E^* =(N^*,P^*,R^*)$ and same can be  obtained by solving the following set of equations
 
 \begin{equation}
\label{eq:4}
\Lambda - \beta N P - d N = 0
\end{equation}
\begin{equation}
\label{eq:5}
 Q - \delta_0 P - \delta_1 P N - \delta_2 P R  = 0
\end{equation}
\begin{equation}
\label{eq:6}
q - \alpha_0 R - \alpha P R = 0
\end{equation}

Equation \eqref{eq:4} gives $N= \displaystyle\frac{\Lambda}{(\beta P + d)}$ and the expression of $R$ is obtained from equation \eqref{eq:6} as  $R = \displaystyle\frac{q}{(\alpha_0 + \alpha P)}$. Putting these two values in \eqref{eq:5}, gives the following cubic polynomial \\
\begin{equation}
\label{eq:7}
P^3 + A_1 P^2 + A_2 P + A_3 =0
\end{equation} 
where;
 $A_1 = \displaystyle \frac{[\delta_1 \Lambda \alpha + \delta_2 q \beta + \delta_0 \alpha_0 \beta + \delta_0 \alpha d - Q \alpha \beta]}{\alpha \beta \delta_0}$, 
 
 $A_2 = \displaystyle\frac{[\delta_1 \Lambda \alpha_0 + \delta_2 q d + \delta_0 \alpha_0 d - Q \alpha_0 \beta - Q \alpha d ]}{\alpha \beta \delta_0}$ 
 and $ A_3 = -\displaystyle\frac{\alpha d Q}{\alpha \beta \delta_0}$
 
equation \eqref{eq:7} has a unique positive root if $ A_1$ and $A_2$ are positive. These inequalities hold true under the following two conditions \\
\begin{equation}
\label{eq:8}
(Q \alpha \beta) < (\delta_1 \Lambda \alpha + \delta_2 q \beta + \delta_0 \alpha_0 \beta + \delta_0 \alpha d) 
\end{equation} 
\begin{equation}
\label{eq:9}
(Q \alpha_0 \beta + Q \alpha d) < (\delta_1 \Lambda \alpha_0 + \delta_2 q d + \delta_0 \alpha_0 d)
\end{equation}

\section{Results and Discussions}
\label{RND}
This section presents the stability analysis of the proposed mathematical model. The conditions of stability are established using the Lyapunov technique. An extensive numerical simulation is also carried out to verify the stability conditions.
\subsubsection*{Stability Analysis}

\hspace{-0.5cm}\textbf{Theorem 3.1} The local stability of the equilibrium point $E^*$ is ascertained by the following condition \\
\begin{equation}
\label{eq:10}
\frac{qQ}{2 P^*R^*} > Max. \lbrace (\delta_2 P^*)^2, (\alpha R^*)^2 \rbrace
\end{equation}

\hspace{-0.5cm}\textbf{Theorem 3.2} The global asymptotic stability of the equilibrium point $E^*$ is ascertained by the following condition\\
\begin{equation}
\label{eq:12}
(\delta_0 + \delta_1 N^* + \delta_2 R^*)  > \frac{2}{\alpha_0} Max. \lbrace (\delta_2 \frac{Q}{\delta_0})^2, (\alpha R^*)^2 \rbrace
\end{equation}
The proof of above theorems are given in the appendix. 
\subsection*{\textit{Numerical Simulation}}
In order to gauge the effect of different parameters on the dynamics of the proposed model and to establish the feasibility of stability conditions, numerical experiments are performed. The set of parameter values (given in table (\ref{para})) is chosen such that the conditions of existence ((\ref{eq:8})- (\ref{eq:9})) of the equilibrium point $E^*$ are satisfied. 
\begin{table}[h]
\centering
\begin{minipage}{.5\textwidth}
\centering
\caption[c]{Values of parameters}
\label{para}
\begin{tabular}{|c| |c| |c|}
\hline
Parameter & Value & Source \\\hline
A & 100 & assumed\\
$\beta$ & 0.03 & assumed\\
q & 10 & \cite{shukla2008effect}\\
Q & 2 & assumed\\
d & 0.01 & assumed\\
$\alpha$ & 0.00003 & \cite{shukla2008effect}\\
$\alpha_0$ & 0.2 & \cite{shukla2008effect}\\
$\delta_0$ & 0.2 & assumed\\
$\delta_1$ & 0.0005 & assumed\\
$\delta_2$ & 0.03 & assumed \\ \hline
\end{tabular}
\end{minipage}
\end{table}


\begin{figure}[h]
\includegraphics[scale=0.6]{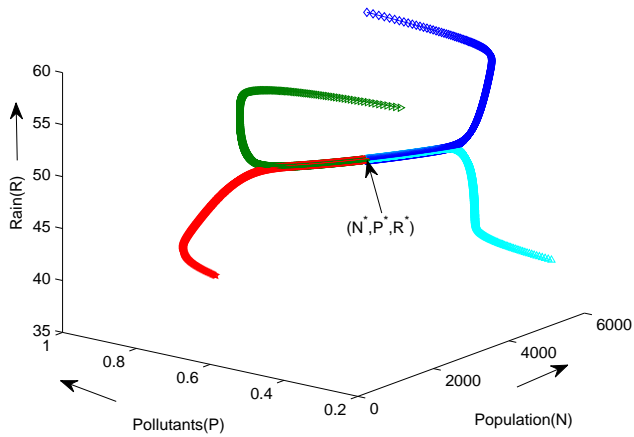}
 \caption{Global stability of $E^*$ in $N-P-R$ plane}
 \label{Fig:pl}
\end{figure}
\begin{figure}[h]
\includegraphics[scale=0.6]{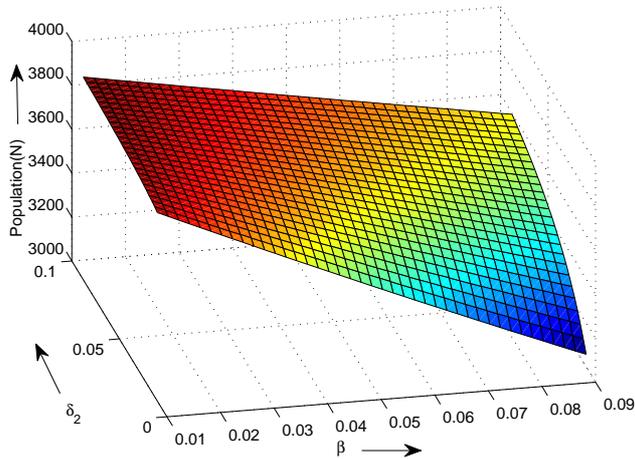}
 \caption{Simultaneous effect of variations in $\beta$ and $\delta_2$ on the human population. }
 \label{Fig:A}
\end{figure}
\begin{figure}[h]
\includegraphics[scale=0.6]{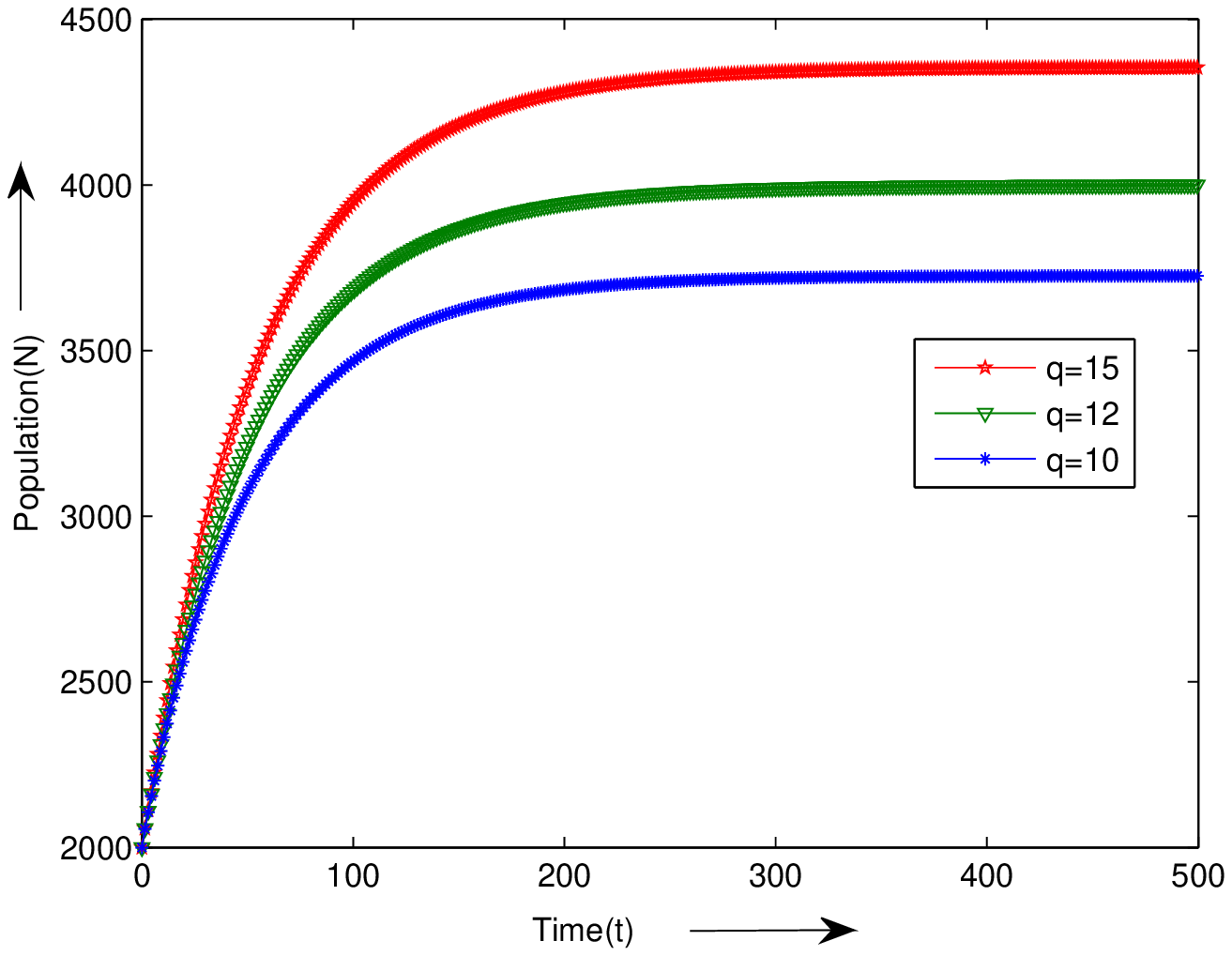}
\caption{The time variation of human population for different values of q }
\label{Fig:B}
\end{figure}
\begin{figure}[h]
\includegraphics[scale=0.6]{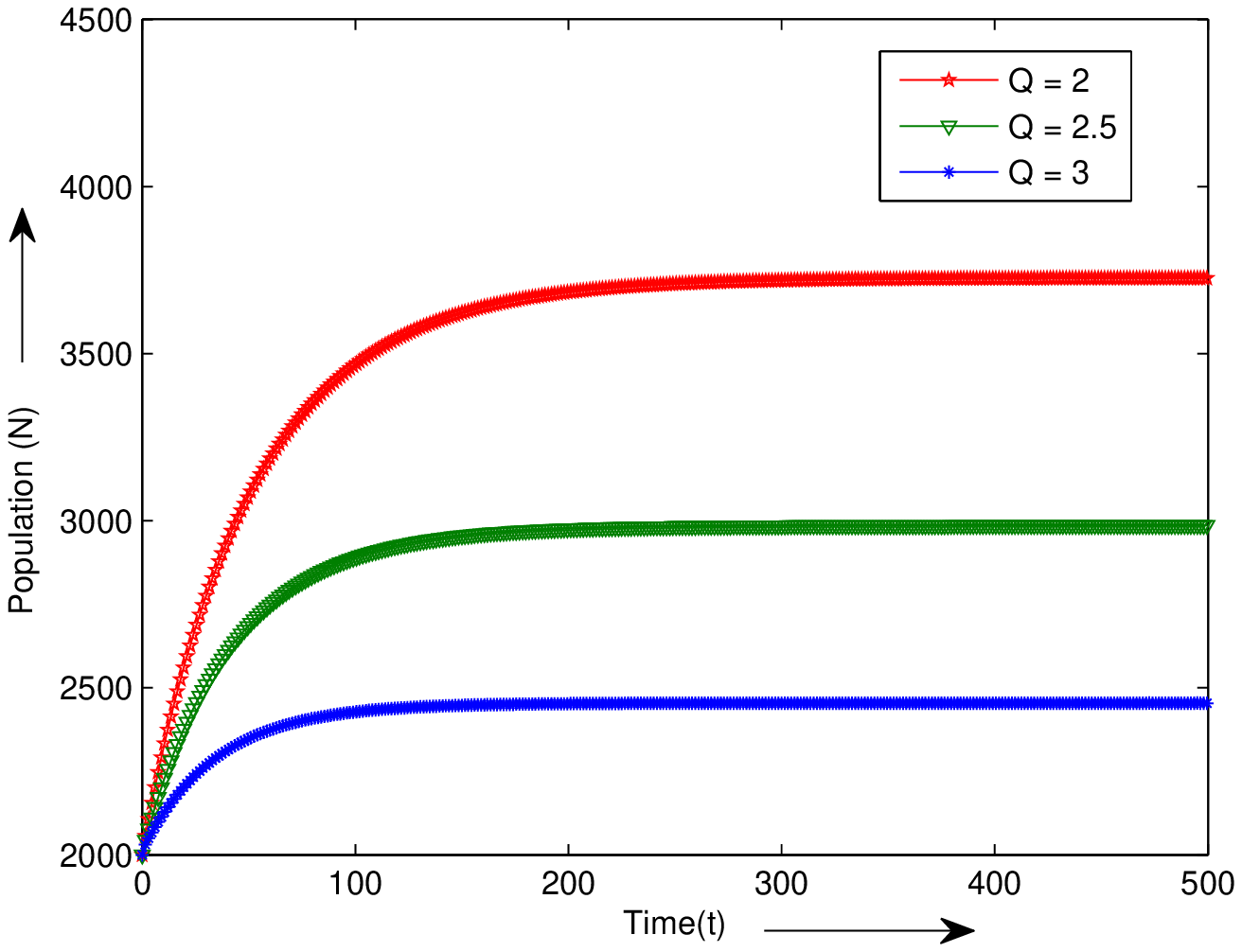}
\caption{The time variation of population for different values of Q }
\label{Fig:C}
\end{figure}
\begin{figure}[h]
\includegraphics[scale=0.6]{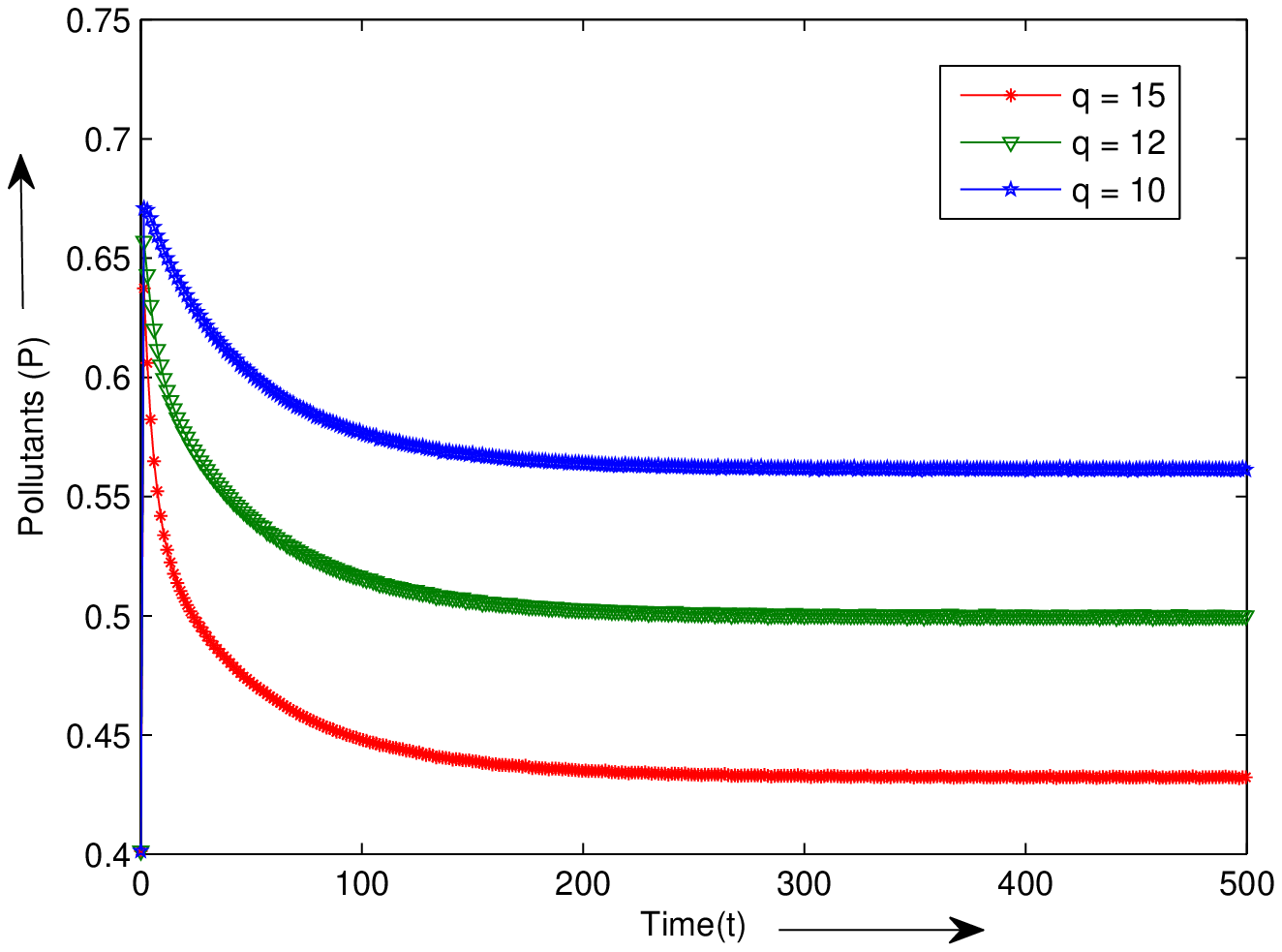}
\caption{The time variation of concentration of air pollutants for different values of q }
\label{Fig:D}
\end{figure}
 The equilibrium components are obtained as: \newline
 
$N^* = 3725.606205, P^* = 0.561375,  R^* = 49.995790$. \newline

The eigenvalues of the jacobian matrix are : \newline
  
    -3.57122702,  -0.01799010 and -0.20000971, \newline
    
    which are all negative. This confirms the local stability of $E^*$. 
 further, for the above values of parameters, conditions for both local and global stability, stated in theorem 3.1 and theorem 3.2, are satisfied. Therefore, selected set of parameter values is valid in the case of our study.
 To examine the stability of the equilibrium point $E^*$, four different trajectories in the $N-P-R$ plane  starting from four different initial conditions are plotted as shown in figure \ref{Fig:pl}. It has been observed that all the trajectories converge to the equilibrium point $E^*$, this confirms the stability of $E^*$ in the $N-P-R$ plane. In figure \ref{Fig:A}, the impact of variation in the air pollutant induced depletion rate of human population $\beta$) and the rate of depletion of air pollutants due to rain ($\delta_2$) on the human population(N) has been studied. From this figure, it is evident that the density of the human population increases with an increase in $\delta_2$. However, the negative effect of the increase in $\beta$ on population is also evident as it decreases with an increase in the value of $\beta$. This shows that rain plays a significant role in the survival of the human population in a polluted environment. To analyze the impact of the change in $q$ (growth rate of rain drops) on the human population, the value of $q$ is increased from 10 to 15 and population curves are plotted, as shown in figure \ref{Fig:B}. It is observed that an increase in the growth rate of rain drops leads to an increase in the population density. This confirms that rain can protect the individuals from the exposure to air pollution. In figure \ref{Fig:C}, numerical simulation is performed to study the effect of air pollutants on the population density. For this purpose, the parameter value of the rate of emission of pollutants, $Q$, is changed from 2 to 3 with an increment of 0.5. The change in human population density versus change in $Q$ is plotted in figure \ref{Fig:C}. It can be observed that as the rate of emission of air pollutants increases in the atmosphere, the population density decreases. Thus, it can be concluded that air pollution causes extreme negative effects on the human population, which results in the decrease of human population density. It is interesting to study the impact of the change in $q$ on the air pollutants, therefore in figure \ref{Fig:D}, the effect of variation of $q$ on the concentration of air pollutants has been studied. The figure clearly shows that concentration of air pollutants decreases significantly with an increase in the value of $q$. Therefore, the present study concludes that rain produces a clean environment by removing various air pollutants. An extensive numerical simulation done in this section validates our analytical findings. 
 
\section{Conclusion}
\label{con}
In present work, a non-linear deterministic model has been formed and studied to understand the scavenging of air pollutants by rain and the manner in which it can help the human population. From the discussion carried out in section (\ref{intro}), there is no doubt to conclude that air pollutants cause childhood mortality, cancer and several other diseases related to the respiratory system and nervous system. Therefore, air pollution is a cause of concern to the society. Rapid industrialization, urbanization, and transportation are the need of modern times, hence these events can not be avoided. This will increase the requirement and need of some effective policies and programs to cope up with the growing problem of air pollution. 

Through different experimental studies \citep{sharma1983atmospheric,pandey1992air,davies1976precipitation}, it has been found that rain plays an active role in the removal of air pollutants as the concentration of pollutants is significantly reduced during the rainy season. Till date, no mathematical attempt has been carried out to study the impact of the removal of air pollutants on the human population density.

Present study demonstrates that rain is an effective and reliable method which can significantly reduce the concentration of  air pollutants and protect living species and environment. 
However, rain is a natural phenomenon and beyond human control. Therefore, agencies are working on artificial rain and different techniques, like mixing of aerosols (cloud seeding) \citep{shukla2010artificial} and electrification of atmosphere \citep{doshi2015feasibility}, are used to produce artificial rain. In the first method, aerosol particles are introduced into the atmosphere by using small air-crafts. This method plays an effective role if water vapor is present in the atmosphere. In such cases, aerosol particles stimulate the formation of cloud droplets and increase the rainfall. The aerosols based technique has recently used by China to produce artificial rain in Beijing \citep{shukla2010artificial}. In the electrification technique, the atmosphere is ionized artificially to produce rainfall. In this technique, emission of chemical using air-crafts is not needed and corona discharge is used to ionize the atmosphere artificially \citep{doshi2015feasibility}. Therefore, use of any of the aforementioned techniques can save the human population from hazardous effects of air pollution.

\section{Appendix 1}
\subsection{Proof of theorem 3.1}
 The jacobian matrix of system \eqref{eq:1}-\eqref{eq:3} around $E^*$ is
\[ P =
\begin{bmatrix}
-(\beta P^* +d ) & -\beta P^* & 0 \\
\delta_1 P^* & -(\delta_0 + \delta_1 N^* + \delta_2 R^*) & -\delta_2 P^* \\
0 & -\alpha R^* & - (\alpha_0 + \alpha P^*) \\
\end{bmatrix}.
\]
Now, consider the following positive definite function \\
\begin{equation*}
W= \frac{1}{2}K_0 N_{1}^2 + \frac{1}{2}K_1 P_{1}^2 + \frac{1}{2} K_2 R_{1}^2.
\end{equation*}
Here $K_i s$ are the positive constants which are chosen appropriately and $N_1, P_1$ and $R_1$ are the small perturbations around $E^*$ i.e $N = N^* + N_1, P = P^* + P_1$ and $ R = R^* + R_1$. Next, the time derivative of $W$ yields \\
\begin{equation*}
\displaystyle\frac{dW}{dt} = k_0 N_1 N_1^{'} + K_1 P_1 P_1^{'} + K_2 R_1 R_1^{'}.
\end{equation*}
Using the above matrix we get \\
\begin{equation}
\begin{array}{ll}
\displaystyle \frac{dW}{dt}= &K_0 \frac{\Lambda}{N^*}N_1^2 - K_1 \frac{Q}{P^*}P_1^2 - K_2 \frac{q}{R^*}R_1^2 - K_0 \beta P^* P_1 N_1 \\

&- K_1 \delta_1 P^* N_1 P_1 - K_1 \delta_2 P^* P_1 R_1 - K_2 \alpha R^* P_1 R_1.
\end{array}
\end{equation}
Now under the following conditions, $W$ is negative definite  \\
\begin{equation*}
K_0 (\beta P^*)^2 < \frac{K_1 Q \Lambda}{2 P^* N^*}, 
\end{equation*}
\begin{equation*}
K_1 (\delta_1 P^*)^2 < \frac{K_0 Q \Lambda}{2 N^* P^*} ,
\end{equation*}
\begin{equation*}
K_1 (\delta_2 P^*)^2  < \frac{K_2 q Q}{2 P^* R^*} ,
\end{equation*}
\begin{equation*}
K_2 (\alpha R^*)^2 < \frac{K_1 q Q }{2 P^* R^*}.
\end{equation*}
If we choose $K_1 = K_2 =1$ and $\displaystyle 4(\beta P^*)^2 (\delta_1 P^*)^2 < K_0 < \frac{(\Lambda Q)^2}{(N^*P^*)^2} $, Then the inequality stated in the theorem is satisfied.

\section{Appendix 2}
\subsection{Proof of theorem 3.2}
Consider the following positive definite function \\
\begin{equation*}
V = \frac{1}{2}K_0(N-N^*)^2 + \frac{1}{2}K_1 (P-P^*)^2 + \frac{1}{2} K_2 (R-R^*)^2.
\end{equation*}
After differentiating and performing some algebraic manipulations we get
\begin{equation*}
\begin{array}{ll}
\displaystyle \frac{dV}{dt} =  - K_0 \beta P (N- N^*)^2 - K_0 d (N- N^*)^2 \\
 - K_1 (\delta_0 + \delta_1 N^* + \delta_2 R^*)(P-P^*)^2 - K_2 \alpha_0 (R-R^*)^2 \\

    - K_2 \alpha P (R-R^*)^2 - K_0 \beta N^* (P-P^*)(N-N^*) \\ 
  
    - K_1 \delta_1 P (N-N^*) (P-P^*) - K_1 \delta_2 P (R-R^*)(P-P^*) \\
   - K_2 \alpha R^* (P-P^*)(R-R^*).
  \end{array}
\end{equation*}
Now, Now under the following conditions, $V$ is negative definite 
\begin{equation*}
K_0 (\beta N^*)^2 < \frac{K_{1} d}{2}(\delta_0 + \delta_{1} N^* + \delta_{2} R^*),
\end{equation*} 
\begin{equation*}
K_1 (\delta_{1} P)^2 < \frac{K_{0} d}{2}(\delta_0 + \delta_{1} N^* + \delta_{2} R^* ) ,
\end{equation*}
\begin{equation*}
K_1 (\delta_{2} P)^2 < \frac{K_2 \alpha_0}{2} (\delta_0 + \delta_1 N^* + \delta_2 R^* ), 
\end{equation*}
\begin{equation*}
K_2 (\alpha R^*)^2 < \frac{K_1 \alpha_0}{2} (\delta_0 + \delta_1 N^* + \delta_2 R^* ). 
\end{equation*}
If we choose $K_1 = K_2 =1$ and $\displaystyle\left(\frac{2 \delta_{1} Q^2}{d \delta_{0}^2}\right) (\delta_0 + \delta_{1} N^* + \delta_{2} R^* )  <K_0 < \left(\displaystyle\frac{d}{2 (\beta N^*)^2}\right) (\delta_0 + \delta_1 N^* + \delta_2 R^* )$, then the condition stated in the theorem is satisfied.

\vspace{2cm}
\footnotesize \textbf{Acknowledgment}  The second author acknowledges the funding provided by IIT Mandi under
the project No. IITM/SG/NTK/008 for carrying out this research. 
\bibliographystyle{apalike} 
\bibliography{AIR}   


\end{document}